\documentclass[prb,preprintnumbers,amsmath,amssymb,floatfix]{revtex4}

\usepackage{graphicx}
\usepackage{subfigure}
\usepackage{dcolumn}

\begin{document}

\title{The design of ultra-strong laser with one-dimensional function photonic crystal}
\author{Xiang-Yao Wu$^{a}$ \footnote{E-mail: wuxy2066@163.com}, Ben-Shan
Wu$^{a}$, Si-Qi Zhang$^{a}$, Xiao-Jing Liu$^{a}$\\ Ji-Ping
Liu$^{a}$, Qing Pan$^{a}$, Ji Ma$^{a}$, Xiao-Ru Zhang$^{a}$, Han
Liu$^{a}$, and Hong Li$^{a}$}
 \affiliation{a. Institute of Physics, Jilin Normal
University, Siping 136000 China}

\begin{abstract}

With the optical kerr effect, the conventional photonic crystal
can be turned into the function photonic crystal under the action
of pump light. In the paper, we have designed the ultra-strong
light source and laser with one-dimensional function photonic
crystal. When the incident light is the ordinary light, the output
is ultra-strong light source, and when the incident light is the
low power laser, the output is ultra-strong laser, the maximum
magnification can be reached $10^{80}$ and even more. Otherwise,
we analyzed the effect of period number, medium refractive index
and thickness, incident angle, the pump light irradiation way and
pump light intensity on the magnification, these results shall
help to optimal design ultra-strong light source and laser. \\

\vskip 10pt


\end{abstract}

\vskip 10pt \maketitle {\bf 1. Introduction} \vskip 10pt

Photonic crystals (PC) are a new kind of materials which
facilitate the control of the light [1-5]. An important feature of
the photonic crystals is that there are allowed and forbidden
ranges of frequencies at which light propagates in the direction
of index periodicity [6-8]. Due to the forbidden frequency range,
known as photonic band gap (PBG) [9-12], which forbids the
radiation propagation in a specific range of frequencies. The
existence of PBGs will lead to many interesting phenomena, e.g.,
modification of spontaneous emission [13-15] and photon
localization [16-18]. Thus numerous applications of photonic
crystal have been proposed in improving the performance of
optoelectronic and microwave devices such as high-efficiency
semiconductor lasers, light emitting diodes, wave guides, optical
filters, high-Q resonators, antennas, frequency-selective surface,
optical limiters and amplifiers [19-21]. These applications would
be significantly enhanced if the band structure of the photonic
crystal could be tuned.

In Refs. [22-25], we have proposed the one-dimensional function
photonic crystal, which is constituted by two media $A$ and $B$,
their refractive indices are the functions of space position.
Unlike conventional photonic crystal (PCs), which is constituted
by the constant refractive indices media $A$ and $B$. We have
studied the transmissivity and the electric field distribution
with and without defect layer, and have designed some optical
devices, such as optical amplifier, attenuator, optical diode and
optical triode by the function photonic crystal.

In the paper, we have designed the ultra-strong light source and
laser with one-dimensional function photonic crystal. When the
incident light is the ordinary light, the output is ultra-strong
light source, and when the incident light is the low power laser,
the output is ultra-strong laser, the maximum magnification can be
reached $10^{80}$ and even more. Otherwise, we analyzed the effect
of period number, medium refractive index and thickness, incident
angle, the pump light irradiation way and pump light intensity on
the magnification, these results shall help to optimal design
ultra-strong light source and laser.

\vskip 10pt {\bf 2. The transmissivity of one-dimensional function
photonic crystal} \vskip 10pt

In Refs. [22-25], we have given the one-dimensional function
photonic crystal transfer matrices $M_{B}$ and $M_{A}$ of the
media $B$ and $A$ for the $TE$ wave, they are
\begin{eqnarray}
M_{B}=\left(%
\begin{array}{cc}
  \cos\delta_{b} & \frac{-i\sin\delta_{b}}{\sqrt{\frac{\varepsilon_{0}}{\mu_{0}}}n_{b}(b)\cos\theta_{i}^{I}} \\
 -in_{b}(0)\sqrt{\frac{\varepsilon_{0}}{\mu_{o}}}\cos\theta_{t}^{I}\sin\delta_{b}
 & \frac{n_{b}(0)\cos\theta_{t}^{I}\cos\delta_{b}}{n_{b}(b)\cos\theta_{i}^{I}}\\
\end{array}%
\right),
\end{eqnarray}
\begin{eqnarray}
M_{A}=\left(%
\begin{array}{cc}
 \cos\delta_{a} & -\frac{i\sin\delta_{a}}{\sqrt{\frac{\varepsilon_{0}}{\mu_{0}}}n_{a}(a)\cos\theta_{i}^{II}} \\
 -in_{a}(0)\sqrt{\frac{\varepsilon_{0}}{\mu_{o}}}\cos\theta_{t}^{II}\sin\delta_{a}
 & \frac{n_{a}(0)\cos\theta_{t}^{II}\cos\delta_{a}}{n_{a}(a)\cos\theta_{i}^{II}}\\
\end{array}%
\right),
\end{eqnarray}
where
\begin{eqnarray}
\delta_{b}=\frac{\omega}{c}n_{b}(b)\cos\theta^{I}_{i}\cdot b,
\hspace{0.2in}
\delta_{a}=\frac{\omega}{c}n_{a}(a)\cos\theta^{II}_{i}\cdot a,
\end{eqnarray}
\begin{eqnarray}
\cos\theta^{I}_{i}
=\sqrt{1-\frac{n_{0}^{2}}{n_{b}^{2}(b)}\sin^{2}\theta_{i}^{0}},
\hspace{0.2in} \cos\theta^{I}_{t}
=\sqrt{1-\frac{n_{0}^{2}}{n_{b}^{2}(0)}\sin^{2}\theta_{i}^{0}},
\end{eqnarray}
and
\begin{eqnarray}
\cos\theta^{II}_{i}
=\sqrt{1-\frac{n_{0}^{2}}{n_{a}^{2}(a)}\sin^{2}\theta_{i}^{0}},
\hspace{0.2in} \cos\theta^{II}_{t}
=\sqrt{1-\frac{n_{0}^{2}}{n_{a}^{2}(0)}\sin^{2}\theta_{i}^{0}}.
\end{eqnarray}
In one period, the transfer matrix $M$ is
\begin{eqnarray}
&&M=M_{B}\cdot M_{A}\nonumber\\
&&=\left(%
\begin{array}{cc}
  \cos\delta_{b} & \frac{-i\sin\delta_{b}}{\sqrt{\frac{\varepsilon_{0}}{\mu_{0}}}n_{b}(b)\cos\theta_{i}^{I}} \\
 -in_{b}(0)\sqrt{\frac{\varepsilon_{0}}{\mu_{o}}}\cos\theta_{t}^{I}\sin\delta_{b}
 & \frac{n_{b}(0)\cos\theta_{t}^{I}\cos\delta_{b}}{n_{b}(b)\cos\theta_{i}^{I}}\\
\end{array}%
\right) \nonumber\\&&
\left(%
\begin{array}{cc}
   \cos\delta_{a} & \frac{-i\sin\delta_{a}}{\sqrt{\frac{\varepsilon_{0}}{\mu_{0}}}n_{a}(a)\cos\theta_{i}^{II}} \\
 -in_{a}(0)\sqrt{\frac{\varepsilon_{0}}{\mu_{o}}}\cos\theta_{t}^{II}\sin\delta_{a}
 &\frac{n_{a}(0)\cos\theta_{t}^{II}\cos\delta_{a}}{n_{a}(a)\cos\theta_{i}^{II}}\\
\end{array}%
\right).
\end{eqnarray}
Where $n_{b}(0)$, $n_{b}(b)$, $n_{a}(0)$ and $n_{a}(a)$ are the
starting point and endpoint values of refractive indices for the
media $B$ and $A$, $b$ and $a$ are the thickness of media $B$ and
$A$, $\theta_{i}^{0}$ is incident angle, $n_{0}$ is air refractive
index.

For the one-dimensional function photonic crystal of structure
$(BA)^N$, its characteristic equation is
\begin{eqnarray}
\left(%
\begin{array}{c}
  E_{1} \\
  H_{1} \\
\end{array}%
\right)&=&M_{B}M_{A}M_{B}M_{A}\cdot\cdot\cdot M_{B}M_{A}\left(%
\begin{array}{c}
  E_{N+1} \\
  H_{N+1} \\
\end{array}%
\right)
\nonumber\\&=&M\left(%
\begin{array}{c}
  E_{N+1} \\
  H_{N+1} \\
\end{array}%
\right)=\left(%
\begin{array}{c c}
  A &  B \\
 C &  D \\
\end{array}%
\right)
 \left(%
\begin{array}{c}
  E_{N+1} \\
  H_{N+1} \\
\end{array}%
\right),
\end{eqnarray}
with the total transfer matrix $M$, we can obtain the transmission
coefficient $t$, it is
\begin{eqnarray}
t=\frac{E_{N+1}}{E_{in}}=\frac{E_{out}}{E_{in}}=\frac{2\eta_{0}}{A\eta_{0}+B\eta_{0}\eta_{N+1}+C+D\eta_{N+1}},
\end{eqnarray}
and the magnification $\beta$ of ultra-strong light source and
laser $\beta$ are defined as
\begin{eqnarray}
\beta=|t|=|\frac{E_{out}}{E_{in}}|=|\frac{2\eta_{0}}{A\eta_{0}+B\eta_{0}\eta_{N+1}+C+D\eta_{N+1}}|.
\end{eqnarray}
Where $E_1=E_{in}+E_{r}$, $E_{in}$ is the incident electric field,
$E_{r}$ is the reflected electric field, $E_{N+1}=E_{out}$ is the
output electric field and
$\eta_{0}=\eta_{N+1}=\sqrt{\frac{\varepsilon_0}{\mu_0}}\cos\theta
_{i}^0 $.

\begin{figure}[tbp]
\includegraphics[width=12cm, height=4.5cm]{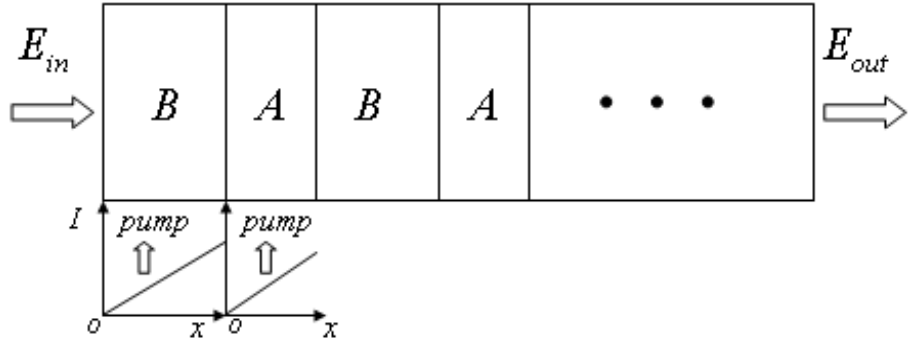}
\caption{The pump light vertically irradiate one-dimensional
photonic crystal, the media $B$ and $A$ are respectively
irradiated by the same pump light.}
\end{figure}

\begin{figure}[tbp]
\includegraphics[width=12cm, height=4.5cm]{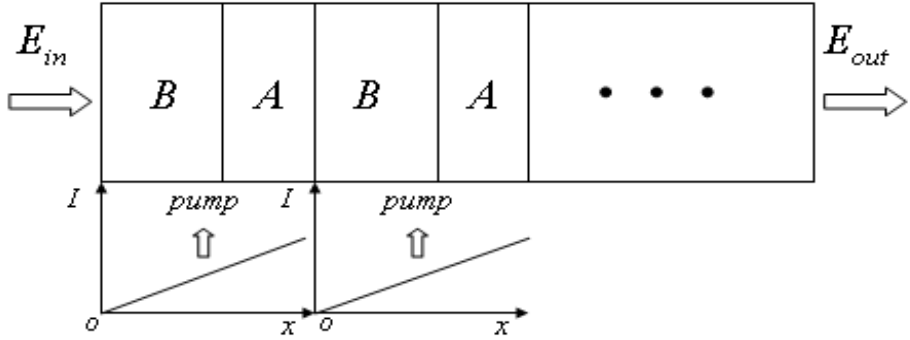}
\caption{The pump light vertically irradiate one-dimensional
photonic crystal, and every one-period $(BA)$ is respectively
irradiated by the same pump light.}
\end{figure}

 \vskip 10pt {\bf 3. The design principle of ultra-strong light source and laser} \vskip
10pt

In the following, we shall explain how to turn the one-dimensional
conventional photonic crystal into the one-dimensional function
photonic crystal, and give the design principle of the
ultra-strong light source and laser. In nonlinear optics, the
medium refractive index is the linear function of light intensity
$I$, which is called the optical kerr effect, it is [26]
\begin{eqnarray}
n(I)=n_0+n_{2}I,
\end{eqnarray}
where $n_0$ represents the usual, weak-field refractive index, the
optical Kerr coefficient $n_{2}=\frac{3}{4n_0^2\epsilon_0
c}\chi^{(3)}$, and $\chi^{(3)}$ is the third-order nonlinear
optical susceptibility.

In Fig. 1, the pump light vertically irradiate one-dimensional
photonic crystal, the media $B$ and $A$ are respectively
irradiated by the same pump light, and the pump light intensity
distribution $I$ is the function of space position $x$, it is
\begin{eqnarray}
I=I_0x,
\end{eqnarray}
substituting Eq. (11) into (10), there is
\begin{eqnarray}
n(x)=n_0+n_{2}I_0x,
\end{eqnarray}
where $I_0$ is the intensity coefficient of pump light. By the the
optical kerr effect, the conventional medium refractive index
$n_0$ become the linear function $n(x)$ of space position $x$,
i.e., the refractive indices $n_b$ and $n_a$ of conventional media
$B$ and $A$ should be become the linear functions $n_b(x)$ and
$n_a(x)$ by the pump light.

With Eq. (12), the refractive indices starting point and endpoint
values of media $B$ and $A$ can be written as:
\begin{eqnarray}
n_{b}(0)=n_{b}, \hspace{0.2in} n_{b}(b)=n_{b}+n_{2b} I_{0}b,
\end{eqnarray}
\begin{eqnarray}
n_{a}(0)=n_{a}, \hspace{0.2in} n_{a}(a)=n_{a}+n_{2a} I_{0}a,
\end{eqnarray}
In Fig. 2, the pump light vertically irradiate one-dimensional
photonic crystal, and every one-period $(BA)$ is respectively
irradiated by the same pump light. In every one-period $(BA)$, the
refractive indices starting point and endpoint values of media $B$
and $A$ are
\begin{eqnarray}
n_{b}(0)=n_{b}, \hspace{0.2in} n_{b}(b)=n_{b}+n_{2b} I_{0}b,
\end{eqnarray}
\begin{eqnarray}
n_{a}(0)=n_{a}+n_{2a} I_{0}b, \hspace{0.2in} n_{a}(a)=n_{a}+n_{2a}
I_{0}(a+b),
\end{eqnarray}
where $n_{b} (n_{a})$ is the refractive index of conventional
medium $B (A)$ (without joining pump light), $n_{b}(0)
(n_{a}(0))$, $n_{b}(b)(n_{a}(a))$ are the refractive indices
starting point and endpoint values of medium $B(A)$ (with joining
pump light), $b (a)$ is the thickness of medium $B (A)$, and
$n_{2b}(n_{2a})$ is the optical Kerr coefficient of medium $B
(A)$.

\vskip 8pt {\bf 4. Numerical result} \vskip 8pt

In this section, we shall calculate the magnification $\beta$ of
the ultra-strong light source and laser designed by the
one-dimensional function photonic crystal. The main parameters
are: The conventional media $B$ and $A$ (without joining pump
light) refractive indices $n_{b}=1.47$, $n_{a}=1.58$, thickness
$b=\frac{\lambda_0}{4n_b}$, $a=\frac{\lambda_0}{4n_a}$, the
optical Kerr coefficient $n_{2b}=2.0\times 10^{-6}$,
$n_{2a}=2.3\times 10^{-6}$, the intensity coefficient of pump
light $I_0=5.0\times10^{10}(W/cm)$, the central wavelength
incident $\lambda_0=1.55\times 10^{-6}m$, the incident angle
$\theta_{i}^{0}=0$, the one-dimensional function photonic crystal
structure is $(BA)^{N}$, where $N$ is the periodic number.
Firstly, we shall calculate the ultra-strong light source and
laser magnification $\beta$ under the pump light action of Fig. 1.
In Fig. 1, the pump light vertically irradiate one-dimensional
photonic crystal, the media $B$ and $A$ are respectively
irradiated by the same pump light. Substituting Eqs. (13) and (14)
into (6), we can obtain the transfer matrix $M$ of one period.
With Eqs. (7), (8) and (9), we can calculate the magnification
$\beta$, they are shown in Figs. 3-11, which give the relation
between magnification $\beta$ and incident light frequency
$\omega$. In Fig. 3, the period number $N=32$, the maximum
magnification $\beta_{max}=1.8\times 10^{14}$. In Fig. 4, the
period number $N=82$, the maximum magnification
$\beta_{max}=3.4\times 10^{36}$. In Fig. 5, the period number
$N=162$, the maximum magnification $\beta_{max}=1.5\times
10^{72}$. From Fig. 3 to Fig. 5, we can find the maximum
magnification $\beta_{max}$ increases with the period number $N$
increasing. In Fig. 6, the intensity coefficient $I_0$ of pump
light is $I_0=2.0\times10^{10}(W/cm)$, other parameters are the
same as In Fig. 5, the maximum magnification
$\beta_{max}=9.6\times 10^{37}$, i.e., the intensity coefficient
$I_0$ of pump light decreases, the maximum magnification
$\beta_{max}$ reduces. In Fig. 7, the medium $B$ thickness
$b=1.6\times\frac{\lambda_0}{4n_b}$, other parameters are the same
as In Fig. 5, the maximum magnification $\beta_{max}=4.2\times
10^{83}$, i.e., the medium thickness increases, the maximum
magnification $\beta_{max}$ increases. In Fig. 8, the medium $B$
refractive index $n_{b}=1.27$, other parameters are the same as In
Fig. 5, the maximum magnification $\beta_{max}=1.3\times 10^{79}$,
i.e., the medium refractive index decreases, the maximum
magnification $\beta_{max}$ reduces. In Fig. 9, the medium $B$
optical Kerr coefficient $n_{2b}=1.6\times 10^{-6}$, other
parameters are the same as In Fig. 5, the maximum magnification
$\beta_{max}=2.3\times 10^{67}$, i.e., the optical Kerr
coefficient decreases, the maximum magnification $\beta_{max}$
reduces. In Figs. 10 and 11, the incident angle are
$\theta_{i}^{0}=\frac{\pi}{6}$ and $\theta_{i}^{0}=\frac{\pi}{3}$,
other parameters are the same as In Fig. 5, the maximum
magnification are $\beta_{max}=4.6\times 10^{75}$ and
$\beta_{max}=1.9\times 10^{84}$, respectively, i.e., with the
incident angle increasing, the maximum magnification $\beta_{max}$
increases. Nextly, we shall calculate the ultra-strong light
source and laser magnification $\beta$ under the pump light action
of Fig. 2. In the Fig. 2, the pump light vertically irradiate
one-dimensional photonic crystal, and every one-period $(BA)$ is
respectively irradiated by the same pump light. Substituting Eqs.
(15) and (16) into (6), we can obtain the transfer matrix $M$ of
one period. With Eqs. (7), (8) and (9), we can calculate the
magnification $\beta$, it is shown in Fig. 12. In Fig. 12, the all
parameters are the same as In Fig. 5, the maximum magnification
are $\beta_{max}=6.8\times 10^{42}$. Comparing Fig. 5 with Fig.
10, the maximum magnification $\beta_{max}$ is different for the
different irradiation way of pump light (Figs. 1 and 2).
Obviously, the irradiation way of Fig. 1 can obtain the more
magnification. By computation above, we can find when the incident
light is the ordinary light or low power laser, the output
ultra-strong light source or laser can be produced by the
one-dimensional function photonic crystal.

The reason of ultra-strong light amplification is the incident
light absorb a lot of energy of pump light. We can compare the
ultra-strong light amplifier with the electric current amplifier.
The incident and output light fields are the equal of the input
and output alternating electrical signals, and the pump light
amount to direct-current power source.

\vskip 10pt {\bf 5. Conclusion} \vskip 10pt

In the paper, we have designed the ultra-strong light source and
laser with one-dimensional function photonic crystal. When the
incident light is the ordinary light, the output light is
ultra-strong light source, and when the incident light is the low
power laser, the output light is ultra-strong laser, the maximum
magnification can be reached $10^{80}$ and even more. Otherwise,
we found the maximum magnification should be increased when the
period number, medium refractive index and thickness, incident
angle and optical Kerr coefficient increase, the maximum
magnification of Fig. 1 is larger than Fig. 2, these results shall
help to optimal design ultra-strong light source and laser.

\vskip 12pt {\bf 6.  Acknowledgment} \vskip 12pt

This work was supported by the Scientific and Technological
Development Foundation of Jilin Province (no.20130101031JC).

\newpage

\begin{figure}[tbp]
\includegraphics[width=16cm, height=8cm]{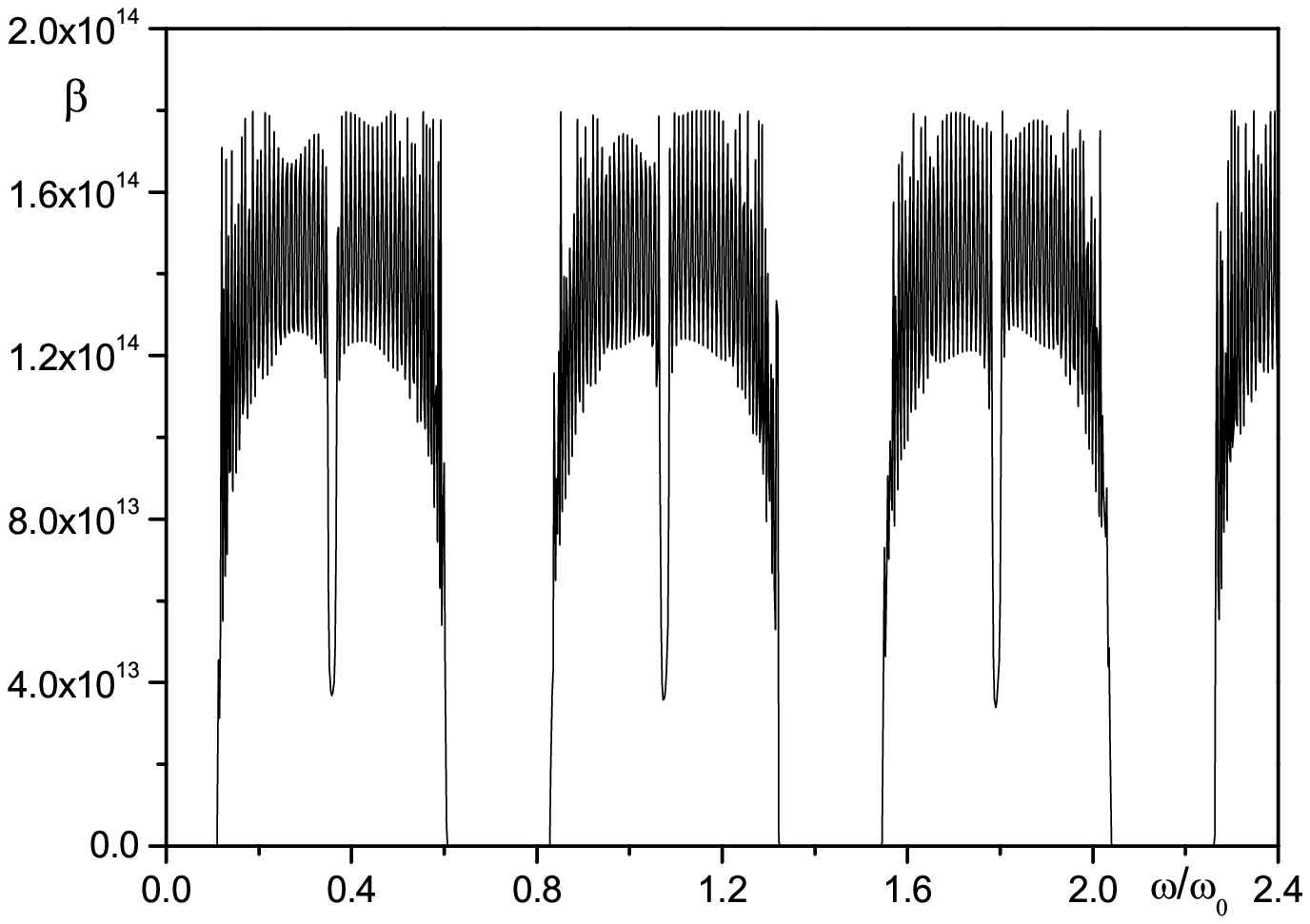}
\caption{The magnification of period number $N=32$.}
\end{figure}

\begin{figure}[tbp]
\includegraphics[width=16cm, height=8cm]{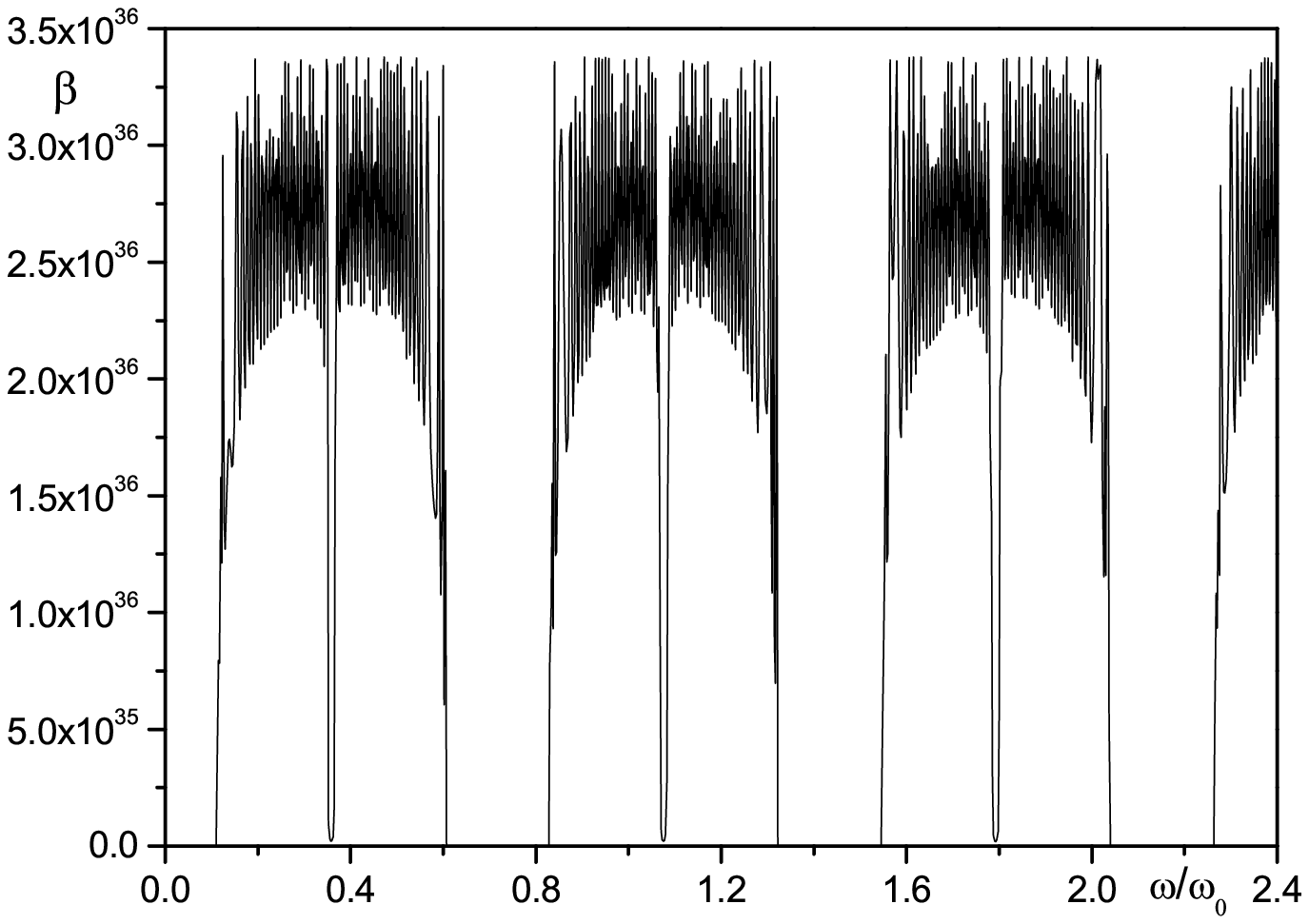}
\caption{The magnification of period number $N=82$.}
\end{figure}

\begin{figure}[tbp]
\includegraphics[width=16cm, height=8cm]{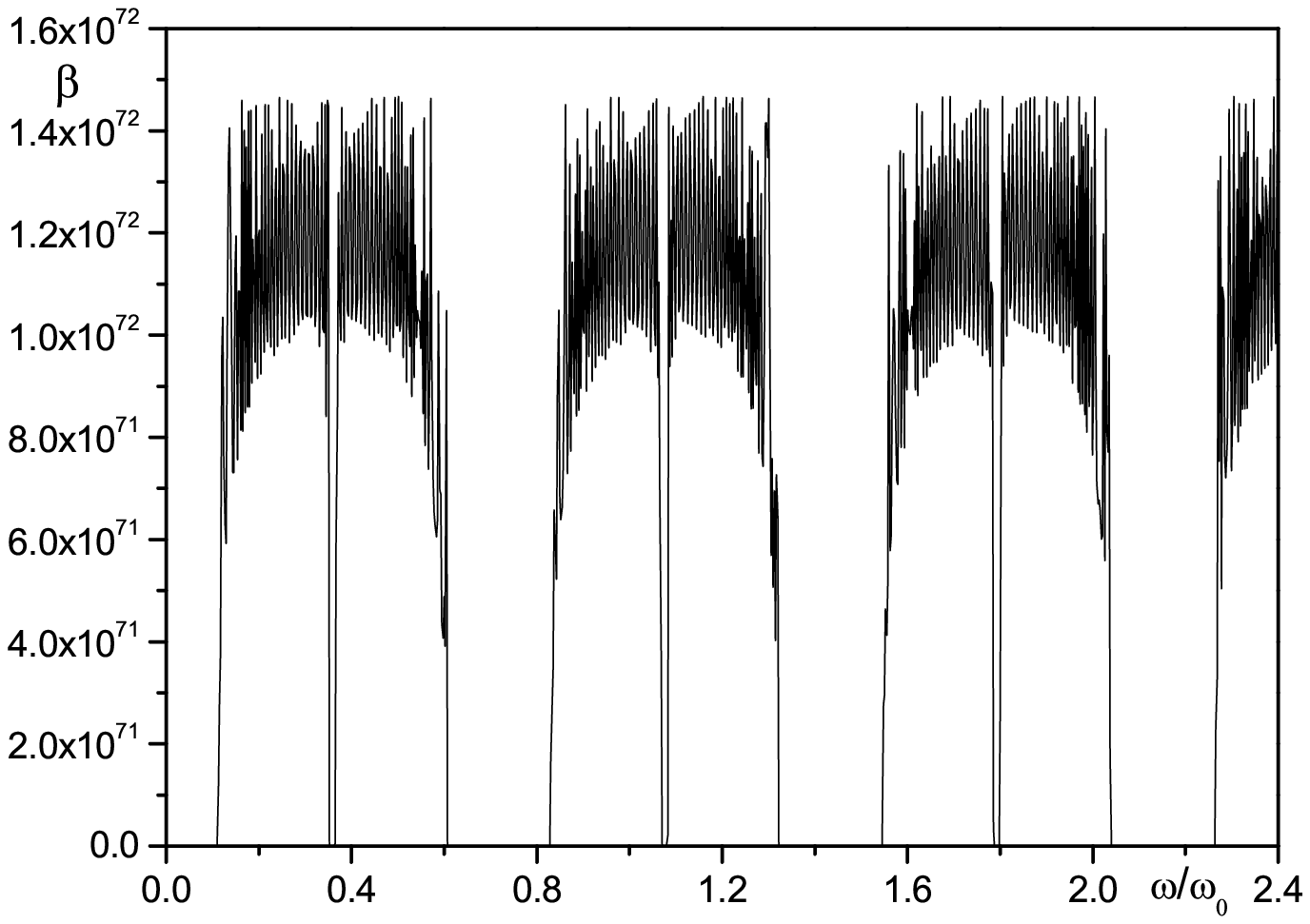}
\caption{The magnification of period number $N=162$.}
\end{figure}

\begin{figure}[tbp]
\includegraphics[width=16cm, height=8cm]{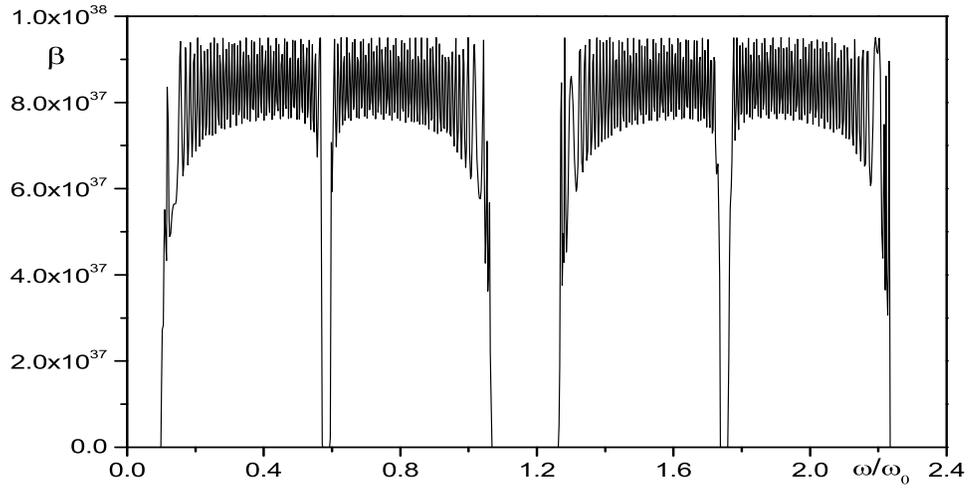}
\caption{The magnification of period number $N=162$ and the
intensity coefficient $I_0=2.0\times10^{10}(W/cm)$.}
\end{figure}

\begin{figure}[tbp]
\includegraphics[width=16cm, height=8cm]{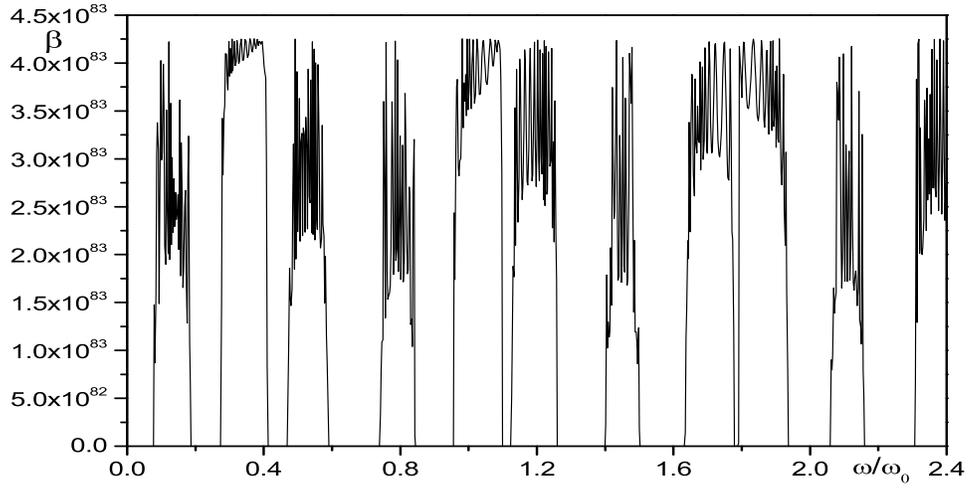}
\caption{The magnification of period number $N=162$ and thickness
$b=1.6\times\frac{\lambda_0}{4n_b}$.}
\end{figure}

\begin{figure}[tbp]
\includegraphics[width=16cm, height=8cm]{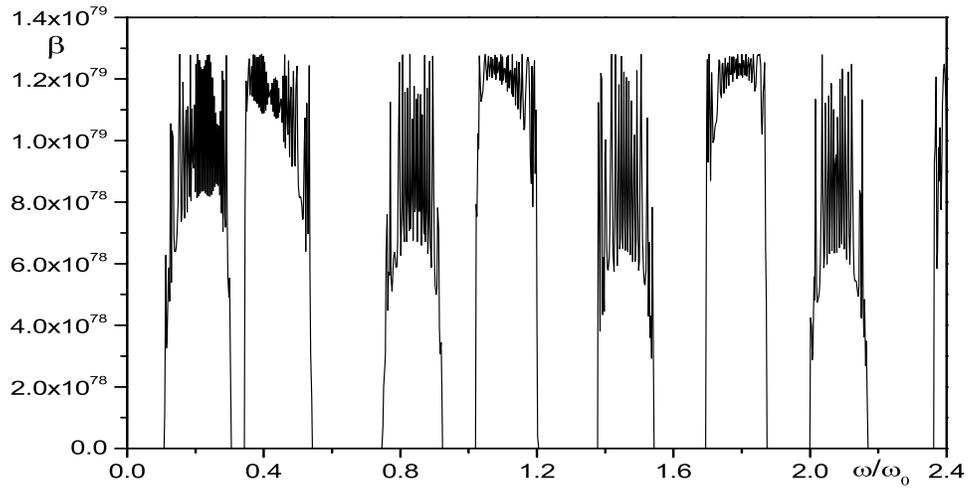}
\caption{The magnification of period number is $N=162$ and the
refractive index $n_{b}=1.27$.}
\end{figure}

\begin{figure}[tbp]
\includegraphics[width=16cm, height=8cm]{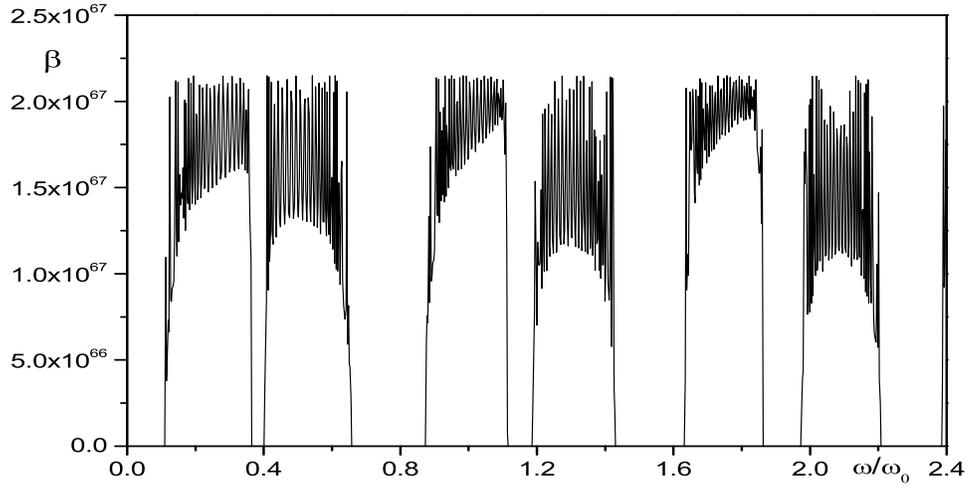}
\caption{The magnification of period number is $N=162$ and the
optical Kerr coefficient $n_{2b}=1.6\times 10^{-6}$.}
\end{figure}

\begin{figure}[tbp]
\includegraphics[width=16cm, height=8cm]{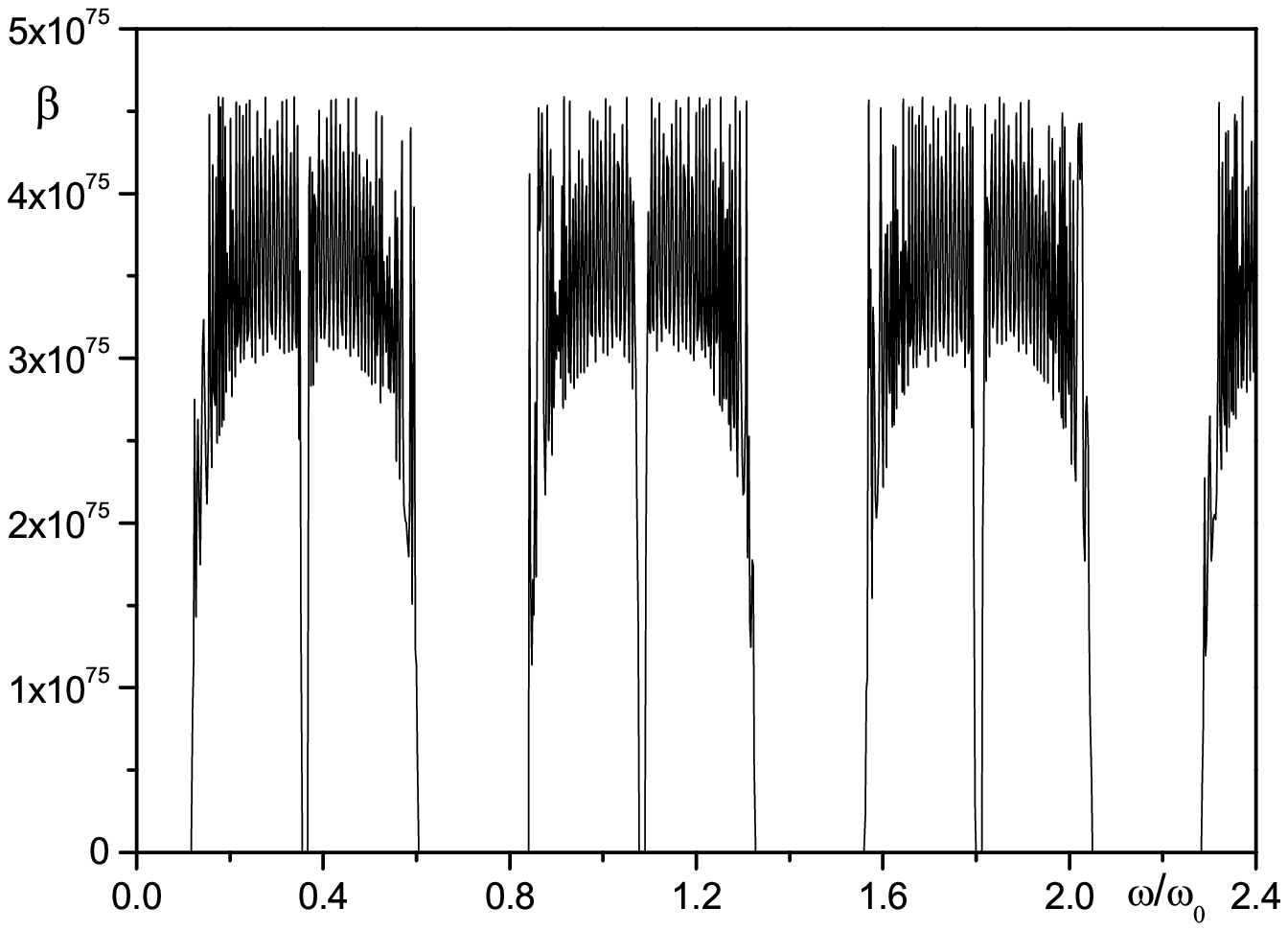}
\caption{The magnification of period number $N=162$, the incident
angle $\theta=\frac{\pi}{6}$.}
\end{figure}

\begin{figure}[tbp]
\includegraphics[width=16cm, height=8cm]{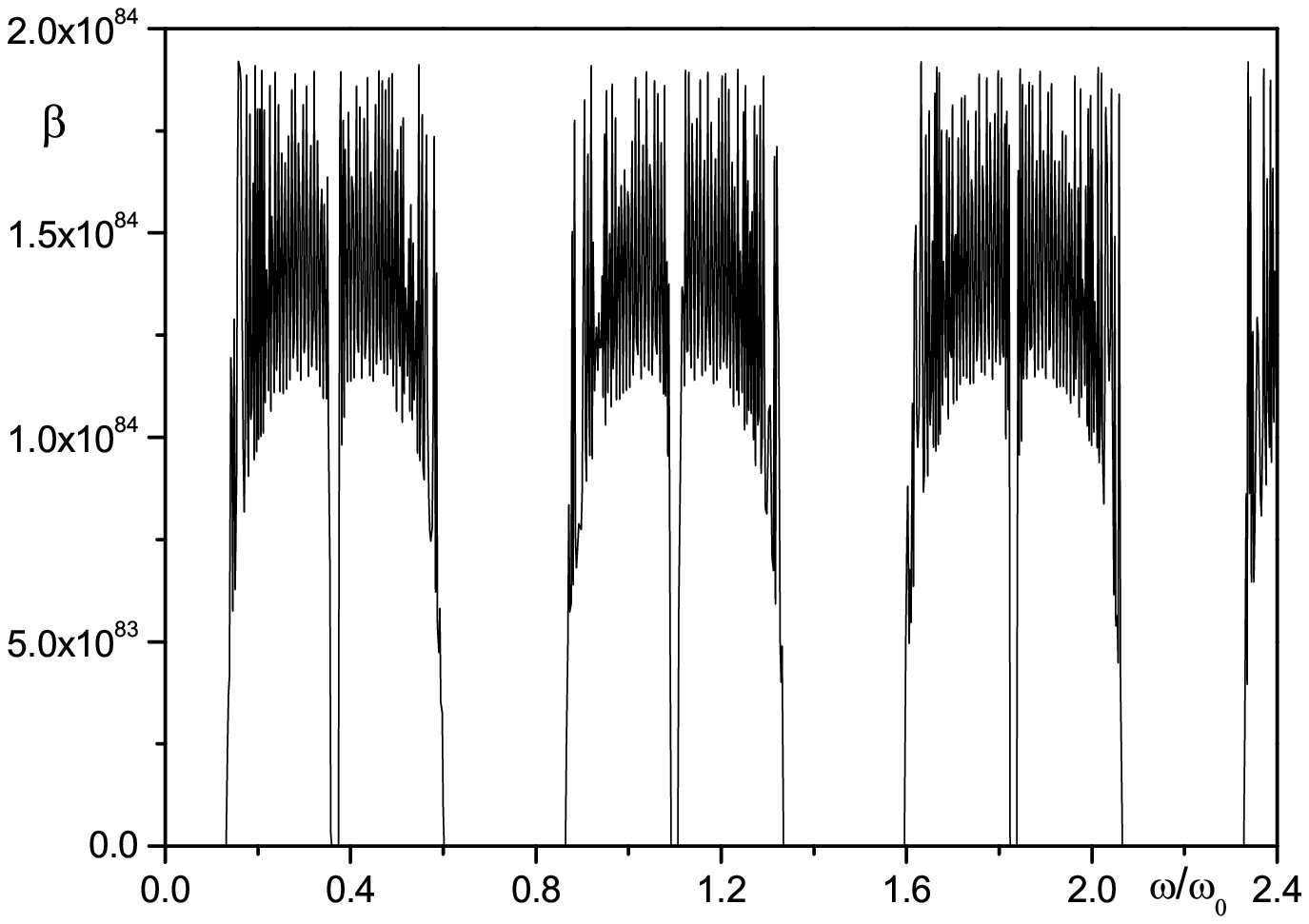}
\caption{The magnification of period number $N=162$, the incident
angle $\theta=\frac{\pi}{3}$.}
\end{figure}

\begin{figure}[tbp]
\includegraphics[width=16cm, height=8cm]{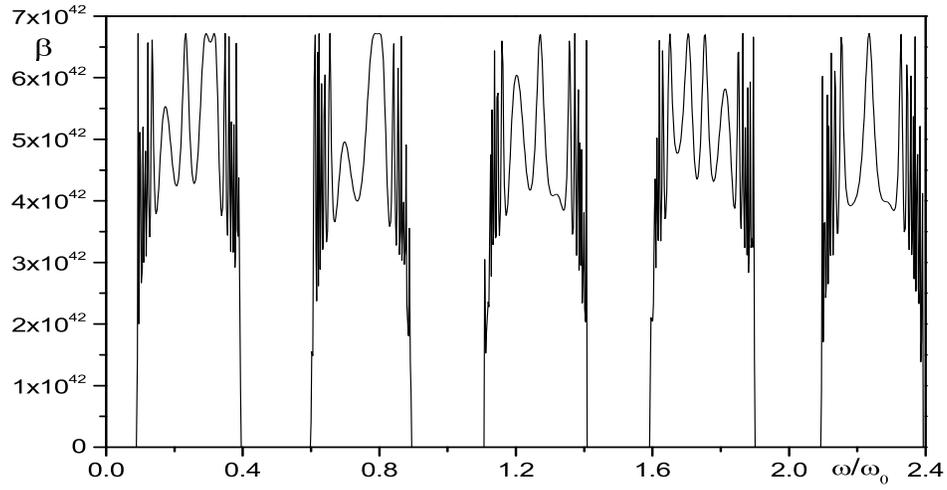}
\caption{The magnification of the all parameters are the same as
In Fig. 5 and the pump light irradiate with Fig. 2.}
\end{figure}

\end{document}